\documentclass[secthm,seceqn,amsthm,ussrhead]{amsart}
\usepackage{amsmath,latexsym}
\usepackage[psamsfonts]{amssymb}
\usepackage{times}
\usepackage[mathcal]{euscript}
\numberwithin{equation}{section} \textwidth=127mm
\newcommand{\bbT}{\mathbb T}
\newcommand{\bbZ}{\mathbb Z}

\renewcommand{\epsilon}{\varepsilon}

\newcommand{\be}{\begin{equation}}
\newcommand{\ee}{\end{equation}}
\newcommand{\no}{\nonumber}

\newcommand{\C}{\mathbb{C}}

\newcommand{\F}{\mathbb{F}}

\newcommand{\N}{\mathbb{N}}

\newcommand{\R}{\mathbb{R}}

\newcommand{\T}{\mathbb{T}}

\newcommand{\Z}{\mathbb{Z}}

\newcommand{\cD}{{\mathcal D}}

\newcommand{\cF}{{\mathcal F}}

\newcommand{\cH}{{\mathcal H}}
\newcommand{\cI}{{\mathcal I}}

\newcommand{\cR}{{\mathcal R}}

\newcommand{\cU}{{\mathcal U}}

\newcommand{\fD}{\mathfrak{D}}

\newcommand{\Img}{\mathop{\mathrm{Im}}}

{\bf}{\it}
\newtheorem{theorem}{Theorem}[section]
\newtheorem{lemma}[theorem]{Lemma}

\newtheorem{hypothesis}[theorem]{Hypothesis}
\newtheorem{remark}[theorem]{Remark}

\date{\today}

\begin{document}
\title[On the structure of the essential spectrum of the three-particle Schr\"{o}dinger...]
{On the structure of the essential spectrum
\\of the three-particle Schr\"{o}dinger operators on a lattice.\\}

 \author{Sergio	 Albeverio$^{1,2,3}$, Saidakhmat  N. Lakaev$^{4}$,
  Zahriddin I. Muminov$^{5}$}

\address{$^1$ Institut f\"{u}r Angewandte Mathematik,
Universit\"{a}t Bonn, Wegelerstr. 6, D-53115 Bonn\ (Germany)}

\address{
$^2$ \ SFB 256, \ Bonn, \ BiBoS, Bielefeld - Bonn;}
\address{
$^3$ \ CERFIM, Locarno and Acc.ARch,USI (Switzerland) E-mail
albeverio@uni.bonn.de}

\address{
{$^4$ Samarkand State University, Samarkand (Uzbekistan)} \
{E-mail:slakaev@mail.ru }}

\address{
{$^5$ Samarkand State University, Samarkand (Uzbekistan)}{
E-mail:zmuminov@mail.ru}}

\begin{abstract} A	system of three
quantum particles  on the three-dimensional lattice $\Z^3$ with
arbitrary "dispersion functions"  having non-compact support and
interacting via short-range pair  potentials is considered.
The energy operators   of the systems of the two-and three-particles
 on the lattice $\Z^3$ in the coordinate and momentum
representations are described as bounded self-adjoint operators
on the corresponding Hilbert spaces. For all sufficiently small nonzero values of the
two-particle quasi-momentum $k\in (-\pi,\pi]^3$ the finiteness of the number
of eigenvalues of the two-particle	discrete  Schr\"odinger operator $h_\alpha(k)$
below the continuous spectrum
 is established.
A location
of
the essential spectrum of
 the three-particle discrete  Schr\"odinger
operator $H(K),K\in (-\pi,\pi]^3$ the
three-particle quasi-momentum, by means of
the spectrum of $h_\alpha(k)$ is described.
It is established that the essential spectrum of $H(K),\,K\in (-\pi,\pi]^3$
 consists of a	finitely many bounded closed intervals.
\end{abstract}
\maketitle



   \section{Introduction}
Location of the essential spectrum of $N$-body Sch\"odinger
operators
for particles moving in $\R^3$
has been extensively studied in many works (  See,
e.g.,\cite{Enss,F,Hunziker,Jorgens,RidSimIV,VanWin ,Zhislin,Zol}
 and
references therein).

We recall that for the three-particle Sch\"odinger operators the
three-particle continuum of the essential spectrum	coincides with
the semi-axis $[0,{ \infty })$. "Two-particle
branches" fill the interval $[{\kappa },{ \infty })$ where ${ \kappa
}\leq 0$ is the lowest eigenvalue of the two-particle Subhamiltonians. Thus,
there are no gaps in the essential spectrum.

%

	In models of solid state physics \cite{GraSch,Mat,Mog,RidSimIV,Yaf3}
	  and also in
lattice	 field theory \cite{LakMin,MalMin} discrete Schr\"{o}dinger operators
are considered, which are lattice analogs of the continuous
three-particle Schr\"{o}dinger operator.

In \cite{L1,L3,LA1,LS1,LA2,LA3,AlbLakzM} for the  Hamiltonians of systems
of three quantum particles moving on the three-dimensional lattice
$\Z^3$ and interacting via zero-range attractive pairs potentials
the location and structure of the essential spectrum has been
investigated.

In particular in \cite{AlbLakzM} it is shown that the essential spectrum of $H(K)$ consists of
no more than four bounded closed intervals. The existence of
infinitely many eigenvalues of $H(0)$ is proven.
It is found that for the number $N(0,z)$ of eigenvalues of $H(0)$ lying below
$z<0$ the following limit exists
$$
\lim _{z\to 0-} \frac {N(0,z)}{\mid \log\mid z\mid\mid }=\cU_0
$$
with $\cU_0>0$. Moreover, for all sufficiently small nonzero values of the
three-particle quasi-momentum $K$ the finiteness of the number $
N(K,\tau_{ess}(K))$ of eigenvalues of $H(K)$ below the essential spectrum
 is established and the asymptotics for the number $N(K,0)$
of eigenvalues	lying below zero is given.


The fundamental difference between the	discrete and continuous
multiparticle \\Schr\"odinger operators is that in the discrete
case the kinetic  energy operator	is not rotationally invariant.
The absence of this property impedes the use of the technique of
separation of variables
(which was essential for considering non-interacting clusters  of
particles in the continuum case).
 In lattice terms the "center-of-mass separation"
corresponds to a realization of	 Hamiltonian as a "fibered
operator", that is, as the "direct integral of a family of
operators" $H(K)$ depending on the values of the total
quasi-momentum $ K{\in} {\bbT}^3=(-\pi,\pi]^3 $	 (see
\cite{GraSch,RidSimIV}). In this case a "bound state" is an
eigenvector of the operator $H(K)$ for some $ K {\in}{\bbT}^3 $.
Typically, this eigenvector depends continuously on $K.$

In the present work we consider the	 system of three
quantum particles  on the three-dimensional lattice $\Z^3$ with
arbitrary "dispersion functions"  having non-compact support and
interacting via short-range pair  potentials.
We describe the energy operators (Hamiltonians)	 for the two-and three-particles
 on the lattice $\Z^3$ in the coordinate and momentum
representations as bounded self-adjoint operators on the corresponding Hilbert spaces.
Then we decompose the energy operators into von Neumann direct integrals
and introduce the two-and three-particle quasimomenta  and relative coordinate systems.
We show that the two-and three-particle fiber operators
$\tilde h_\alpha(k),$ $k \in {\bbT}^3,$ and $\widetilde H(K),$ $K
\in {\bbT}^3,$ are unitarily equivalent to the "two-and three-particle
discrete Schr\"odinger operators $ h_\alpha(k),$ $k \in {\bbT}^3,$
and $H(K),$ $K \in {\bbT}^3,$" on the Hilbert spaces $L_2({\T}^3) $
and $L_2(({\T}^3)^2).$

We also obtain a generalization of the Birman-Schvinger principle for the two-particle
discrete Schr\"odinger operators $h_\alpha(k)$(Theorem 5.4) and, using this,
 we establish that for all sufficiently small values of the
two-particle quasi-momentum $k\in (-\pi,\pi]^3$	 the number
of eigenvalues of  $h_\alpha(k)$ below the continuous spectrum is finite (Theorem 4.1).

We describe a location
of
the essential spectrum for
 the three-particle discrete  Schr\"odinger
operator $H(K)$ by means of
the eigenvalues of the two-particle	  operators $h_\alpha(k)$ (Theorem 4.2).

Our main results is that the essential spectrum of the discrete three-particle	Schr\"odinger
operator $H(K),K\in \T^3$ consists of only a finitely many bounded closed intervals (Theorem 4.3).
Our proof is based on the fact that
  for each
$\alpha=1,2,3$ and some $k \in \T^3$  the operator
$h_\alpha(k)$ has finitely many eigenvalues below the bottom of
the continuous spectrum.

The plan of	 the paper is as follows:

Section 1 is an introduction.

In section 2 the Hamiltonians of systems of two and
three-particles in coordinate and momentum representations are
described as bounded self-adjoint operators in the corresponding
Hilbert spaces.

In section 3 we introduce the total quasi-momentum and decompose
the energy operators into von Neumann direct integrals.
We show, choosing
relative coordinate systems, that the
 "two"-and "three-"particle fiber operators are unitarily equivalent
to the family of bounded self-adjoint operators acting in the
Hilbert spaces $L_2(\T^3)$ and $L_2((\T^3)^2).$

In section 4 we	 state
the main results of the paper.

In section 5 we study spectral properties of the two-particle
discrete Schr\"{o}dinger operators $h_\alpha(k),k\in \T^3,\alpha=1,2,3$ on the
three-dimensional lattice $\Z^3.$

In section 6 we introduce the "channel operators" and prove that
the spectrum of "channel operator" is the  union of a finite
number of bounded closed intervals.

In section 7  applying the	Faddeev	 type system of integral equations we establish the
location of the essential spectrum of $H(K)$ (Theorem \ref{esss}) and  prove the main
result (Theorem \ref{finite}).

Throughout the paper we adopt the following conventions: For each
$\delta>0$ the notation $U_{\delta}(0) =\{K\in {\bbT}^3:|K|<\delta
\}$ stands for a $\delta$-neighborhood of the origin.
 The subscript $\alpha$ (and also $\beta$
and $ \gamma$) always equal to $1$ or $2$ or $3$ and
$\alpha\neq\beta,\beta\neq\gamma,\gamma\neq\alpha.$

\section{Energy operators of systems of two and three arbitrary
particles on the lattice $\Z^3$ in the coordinate and momentum
representations}

Let $\Z^{\nu}$-$\nu$ dimensional lattice.

  The free Hamiltonian
$\widehat H_0$ of a system of
 three quantum mechanical particles on the	three-dimensional
 lattice $\Z^3$ is defined in terms of three functions $
\hat{\varepsilon}_\alpha(\cdot)$ corresponding to the particles
$\alpha=1,2,3$ (called "dispersion functions" in the physical
literature, see,e.g. \cite{Mat}).
 The operator $\widehat H_0$ usually associated with the following bounded
self-adjoint operator on the Hilbert space $\ell_2(({\bbZ}^3)^3)$:
\begin{equation*}
\widehat{H}_0=	\fD_{x_1}+	\fD_{x_2}+ \fD_{x_3},
\end{equation*}
with $ \fD_{x_1}= D_1\otimes I\otimes I$, $
\fD_{x_{2}}=I \otimes  D_2 \otimes I$ and  $
\fD_{x_3}=I\otimes I\otimes	 D_3$, where
$I$ is identity operator in $\ell_2(\Z^3)$ and \begin{equation*}
( D_{\alpha}\hat{\psi})(x)= \sum_{s\in
{\Z}^3}\hat{\varepsilon}_\alpha(s) \hat{\psi}(x+s),\quad
\hat{\psi} \in \ell_2(\Z^3).
\end{equation*}

Here $ \hat{\varepsilon}_\alpha(\cdot), \alpha=1,2,3 $ are assumed
to be real-valued bounded functions having non	compact support in
 $\Z^3$	 and describing the dispersion low of the
 corresponding particles (see, e.g.,\cite{Mat}).

The three-particle Hamiltonian $\widehat H$ of the
quantum-mechanical three-particles
 systems with two-particle pair	 interactions
$\hat v_{\beta\gamma},\,\beta\gamma=12,23,31$  is a bounded
perturbation of the free Hamiltonian $\widehat H_0$
\begin{equation}\label{total}
 \widehat{H}=\widehat{H}_0-\widehat{V}_{1}-\widehat{V}_{2}-
\widehat{V}_{3},
\end{equation}
where $\widehat{V}_{\alpha}, \alpha=1,2,3 $ are multiplication
operators on $\ell_2(({\bbZ}^3)^3)$
\begin{equation*}
(\widehat{V}_{\alpha}\hat{\psi})(x_1,x_2,x_3)=
\hat{v}_{\beta\gamma}(x_\beta-x_\gamma)\hat{\psi}(x_1,x_2,x_3),
\quad \hat{\psi} \in \ell_2(({\Z}^3)^3),
\end{equation*}
and $\hat{v}_{\beta\gamma}$ is	bounded real-valued function.

Throughout this paper we assume the following additional
Hypothesis.

\begin{hypothesis}\label{eps} The functions
$\hat{\varepsilon}_\alpha(s),\,\alpha=1,2,3$   satisfy the
following conditions: $$
\begin{array}{lll}
\text{(i)} \quad \hat{\varepsilon}_{\alpha} (s)\,\,\mbox{ depends
only on}\,\,|s|=|s^{(1)}|+|s^{(2)}|+|s^{(3)}|,
\,s=(s^{(1)},s^{(2)},s^{(3)})\in {\Z}^3;\hspace{0.8cm}\\
\text{(ii)}\quad \mbox{ there exist	 numbers}\,\,a,C>0\,\,\mbox{
such that}\,\, |\hat{\varepsilon}_{\alpha}(s)| \leq C
\exp{(-a|s|)},\,s\in
\Z^3;\\
\mbox{(iii)}\quad \hat \varepsilon_{\alpha}(s)<0,\,|s|=1\quad
\mbox{and}\quad \hat \varepsilon_{\alpha}(s)\leq 0,\,|s|>1,\,s\in
\Z^3.
\end{array}
$$
 \end{hypothesis}

\begin{remark}\label{mas}
The number $
m_{\alpha}=3(-\sum\limits_{s\in {\Z}^3}
{[(s^{(1)})^2+(s^{(2)})^2+(s^{(3)})^2]} \hat
\varepsilon_{\alpha}(s))^{-1}
>0$ means
 the (effective) mass  of the particle $\alpha$.
\end{remark}

\begin{hypothesis}\label{hy} Assume that $\hat v_{\beta \gamma}(s),\,\beta \gamma=12,23,31$
are	 real even
nonnegative
functions on $\Z^3$ and verifying
\begin{equation}\lim _{|s|\to
 \infty}|s|^{3+\kappa}\hat v_{\beta \gamma}(s)=0,\quad \kappa>0. \end{equation}
\end{hypothesis}

It is clear that under Hypothesis \ref{eps} and	 \ref{hy} the
three-particle
 Hamiltonian \eqref{total} is  a bounded self-adjoint operator on the Hilbert space
 $\ell_2(({\bbZ}^3)^3)$.

Similarly as we introduced $\widehat H,$ we shall introduce the
corresponding two-particle Hamiltonians $\hat{h}_\alpha,\,\alpha
=1,2,3$ as bounded self-adjoint operators on the Hilbert space
$\ell_2(({\bbZ}^3)^2)$
\begin{equation*}
\hat{h}_\alpha =\hat{h}_\alpha^0- \hat{v}_{\alpha },
\end{equation*}
where
\begin{equation*}
\hat{h}_\alpha^0 =	\cD_{x_\beta}+	\cD_{x_\gamma},
\end{equation*}
with $ \cD_{x_\beta}= D_\beta \otimes I$,
$\cD_{x_\gamma}=I \otimes  D_\gamma$ and
\begin{equation*}
(\hat{v}_{\alpha}\hat{\varphi})(x_\beta,x_\gamma) =\hat
v_{\beta\gamma}(x_\beta-x_\gamma)
\hat{\varphi}(x_\beta,x_\gamma),\quad \hat{\varphi} \in
\ell_2(({\bbZ}^3)^2).
\end{equation*}

Let us rewrite our operators in the momentum representation.
 Let ${ \cF}_m:L_2(({\bbT}^3)^m) \rightarrow \ell_2((
{\bbZ}^3)^m)$ denote the standard Fourier transform,
 where ${({\T}^3)^m},\,m\in \N$ denotes the Cartesian
$m$-th power of the set ${\bbT}^3=(-\pi,\pi]^3.$

\begin{remark} The operations addition	and multiplication by
real number of elements of ${\bbT}^3\subset\R^3$ should be regarded as operations on
$\R^3$ modulo $(2\pi \Z^1)^3$.
For example, let
  $$a=(\frac{2\pi}{3},\frac{3\pi}{4},\frac{11\pi}{12}),\quad
b=(\frac{2\pi}{3},
\frac{\pi}{2},\frac{5\pi}{6})\in \T^3$$ then
$$
a+b=(-\frac{2\pi}{3},- \frac{3\pi}{4},-\frac{\pi}{4})\in
\T^3,\quad 12 a= (0,\pi,\pi)\in \T^3. $$
\end{remark}

The three-resp. two-particle Hamiltonians (in the momentum
representation) are given by the bounded self-adjoint operators on
the Hilbert spaces
 $L_2(({\T}^3)^3)$ resp. $L_2(({\T}^3)^2)$
 as follows
 \begin{equation*}
 H={\cF}_3^{-1} \widehat H {\cF}_3
 \end{equation*}
resp.
 \begin{equation*}
h_\alpha={\cF}_2^{-1} \hat h_\alpha {\cF}_2,\quad  \alpha=1,2,3.
\end{equation*}
One has
\begin{equation*}
H=H_0-{V}_{1}-V_{2}-V_{3},
\end{equation*}
where
\begin{equation*}
{H}_0= \hat \fD_{k_1}+ \hat \fD_{k_2}+\hat \fD_{k_3},
\end{equation*}
with $\hat \fD_{k_1}=\hat D_1\otimes I\otimes I$, $\hat	 \fD_{k_{2}}=I \otimes
\hat D_2 \otimes I$,   $\hat \fD_{k_3}=I\otimes I\otimes \hat D_3$\\
and $\hat D_\alpha,\,\alpha=1,2,3$ is the multiplication operator by
the function $ \varepsilon _\alpha (k)$
 \begin{equation*}
(\hat D_{\alpha} f)(k)= {\varepsilon}_\alpha(k) f(k),\quad f \in
L_2(\T^3)
\end{equation*}
 and
$V_{\alpha},\alpha=1,2,3$  are integral operators of convolution
type
\begin{align*}
&(V_{\alpha}f)(k_1,k_2,k_3)
\\
& = (2\pi)^{-\frac{3}{2}}{\int\limits_{({\T}^3)^3} }
 v_{\alpha}
\bigg (\frac{k_\beta-k_\gamma-k_\beta'+k_\gamma'}{2}\bigg ) \delta
(k_\alpha -k_\alpha ')\delta (k_\beta +k_\gamma -k_\beta
'-k_\gamma ')f(k'_1,k'_2,k'_3) dk'_1 dk_2'dk'_3,\\ &f \in
L_2(({\T}^3)^3),
\end{align*}
where $\delta (\cdot)$ denotes the	Dirac delta-function at the
origin.

Here the functions $\varepsilon_\alpha(k),v_\alpha (k),\, \alpha=1,2,3
$  are	given by the Fourier series $\cF_1^{-1}$ and
are of the form
\begin{align*}\no
&\varepsilon _\alpha(k)= \sum_{s\in
{{{{\Z}^3}}}}\hat{\varepsilon}_\alpha\,(s)\,e^{i(k,s)}, \quad
v_{\alpha} (k)=(2\pi )^{-3/2}\sum_{s\in {{\Z}}^3}\hat{v}_{\beta
\gamma }\,(s)\,e^{i(k,s)},\no\\
& \beta \gamma=12,23,31,\,\alpha \neq \beta \neq \gamma
\end{align*}
 with
$$
(k,s)={\sum }_{j=1}^3 k^{(j)}s^{(j)}, \quad
k=(k^{(1)},k^{(2)},k^{(3)})\in {\R}^3,\quad
s=(s^{(1)},s^{(2)},s^{(3)})\in {{\Z}}^3.
$$

For the two-particle Hamiltonians $h_\alpha,\alpha=1,2,3$ we have:
\begin{equation*}
h_\alpha =h_\alpha ^0-v_{\alpha},
 \end{equation*}
 where
\begin{equation*}
{h}_\alpha^0 =	\hat\cD_{k_\beta}+ \hat\cD_{k_\gamma},
\end{equation*}
with $ \hat\cD_{k_\beta}=\hat D_\beta \otimes I$, $\hat \cD_{x_\gamma}=I
\otimes \hat D_\gamma$
and
\begin{equation}\no\label{interaction}
(v_{\alpha}f)(k_\beta ,k_\gamma )=
(2\pi)^{-\frac{3}{2}}{\int\limits_{({\T}^3)^2} }
 v_{\alpha}
\bigg (\frac{k_\beta-k_\gamma-k_\beta'+k_\gamma'}{2}\bigg )
 \delta (k_\beta
+k_\gamma -k_\beta '-k_\gamma ')f(k_\beta ',k_\gamma')dk_\beta
'dk_\gamma',
\end{equation}
$ f\in L_2(({\T}^3)^2).$

 \section{Decomposition of the energy operators into von Neumann direct integrals.
 Quasimomentum and coordinate systems}

 Given $m\in \N$, denote by $\hat U^m_s$, $s\in {\Z}^3$ the unitary operators on the Hilbert space
  $\ell_2(({\Z}^3)^m)$ defined as:
\begin{equation*}
(\hat U^m_sf)(n_1,n_2,..., n_m)=f(n_1+s,n_2+s,...,n_m+s),\quad
f\in \ell_2(({\Z}^3)^m).
\end{equation*}
We	easily see that
\begin{equation*}
  \hat U^m_{s+p}=\hat U^m_s\hat U^m_p,\quad s,p \in \Z^3,
  \end{equation*}
that is, $\hat U^m_s,s\in\Z^3 $ is a unitary  representation of
the abelian group $\Z^3.$

Via the Fourier transform $ \cF_m$ the unitary representation of
$\Z^3$ in  $\ell_2(({\Z}^3)^m)$ induces a representation of the
group $\Z^3$ in the Hilbert space $L_2(({\T}^3)^m)$	 by unitary
(multiplication) operators $U^m_s= \cF_m^{-1}\hat U^m_s\cF_m$,
$s\in \Z^3$ given by:
\begin{equation}\label{grup}
(U_s^mf)(k_1,k_2,...,k_m)= \exp \big (-i(s,k_1+k_2+...+k_m)\big
)f(k_1,k_2,...,k_m),
\end{equation}
\begin{equation*}
f\in L_2((\T^3)^m).
\end{equation*}

Decomposing the Hilbert space $ L_2((\T^3)^m)$	into the direct
integral
\begin{equation*}
L_2((\T^3)^m)= \int_{K\in {\T}^3} \oplus L_2(\F_K^m)d K,
\end{equation*}
where
\begin{equation*}
\F_K^m=\{(k_1,k_2,..., k_m){\ \in }({\T}^3)^m:k_1+k_2+...+k_m = K
\},\quad K\in {\T}^3,
\end{equation*}
we obtain the corresponding decomposition of the unitary
representation $U_s^m$,
 $s \in \Z^3$ into the	direct integral
\begin{equation*}
U_s^m= \int_{K\in {\T}^3} \oplus U_s(K)d K,
\end{equation*}
where
\begin{equation*}
U_s(K)=\exp(-i(s,K)) I
\quad \text{on} \quad L_2(\F_K^m)
\end{equation*}
and $I=I_{L_2(\F_K^m)}$ denotes the identity operator on the
Hilbert space $ L_2(\F_K^m)$.

The above Hamiltonians $\widehat H$ and $\hat h_\alpha,\,
\alpha=1,2,3$ obviously commute with the groups of translations
$\hat U^3_s$  and $\hat U^2_s$, $s\in \Z^3$, respectively, that
is,
\begin{equation*}
\hat U^3_s\widehat H=\widehat H\hat U^3_s, \quad s\in \Z^3
\end{equation*}
and
\begin{equation*}
\hat U^2_s\hat h_\alpha=\hat h_\alpha\hat  U^2_s, \quad s\in \Z^3,
\quad \alpha=1,2,3.
\end{equation*}
Correspondingly, the Hamiltonians $ H$ and $ h_\alpha,\,
\alpha=1,2,3$ (in the momentum representation) commute with the
groups	$ U^m_s$, $s\in \Z^3$ given by \eqref{grup} for $m=3$ and
$m=2,$ respectively.

  Hence, the operators
  $H$ and $h_\alpha,\,\,
\alpha=1,2,3,$ can be decomposed into the direct integrals
\begin{equation*}
H=\int\limits_ {K \in {\T}^3}\oplus \widetilde H(K)dK \quad
\mbox{and}\quad h_\alpha= \int\limits_{k \in {\T}^3}\oplus\tilde
h_\alpha(k)d k, \quad \alpha=1,2,3,
\end{equation*}
with respect to	 the decompositions
\begin{equation*}
L_2 (( {\T}^3)^3) = \int\limits_ {K \in {\T}^3} {\ \oplus } L_2 (
\F_K^3 ) dK \quad \text{and}\quad L_2 (( {\T}^3)^2) = \int\limits_
{k \in {\T}^3} {\ \oplus } L_2 ( \F_k^2 ) d k,
\end{equation*}
respectively.

For any permutation $\alpha\beta\gamma$ of $123$
 we set:
\begin{equation}\no\label{mass}
l_{\beta\gamma}\equiv\frac{m_\gamma}{m_\beta+m_\gamma},\quad
M\equiv\sum_{\alpha=1}^3 m_\alpha,\quad l_{\alpha}\equiv
\frac{m_\alpha}{ M},
\end{equation}
where the quantity $m_\alpha$ entered in Remark	 \ref{mas}.

Given a cyclic permutation ${\alpha}{\beta}{\gamma}$  of $123$ we
introduce the mappings
\begin{equation*}
\pi^{(3)}_\alpha:(\T^3)^3\to (\T^3)^2,\quad
\pi^{(3)}_\alpha((k_\alpha, k_\beta, k_\gamma))=(q_\alpha,
p_\alpha)
\end{equation*}
and
\begin{equation*}
\pi^{(2)}_\alpha:(\T^3)^2\to \T^3,\quad
\pi^{(2)}_\alpha((k_\beta, k_\gamma))=q_\alpha,
\end{equation*}
 where
\begin{align*}
&q_\alpha= l_{\beta\gamma} k_\beta-l_{\gamma\beta} k_\gamma
 \quad \text{ and } \quad
 p_\alpha= l_\alpha (k_\beta+k_\gamma)-(l_\beta+l_\gamma)k_\alpha
 .
\end{align*}

Denote by $\pi^{(3)}_{K}$ , $K\in \T^3$ resp. $\pi^{(2)}_k$, $k\in
\T^3$ the restriction of $\pi^{(3)}_\alpha$resp.$\pi^{(2)}_\alpha$
onto $\F_K^3\subset (\T^3)^3$ resp. $\F_k^2\subset (\T^3)^2$, that
is,
\begin{equation}\label{project}
\pi^{(3)}_{K}=\pi^{(3)}_\alpha\vert_{\F_K^3}\quad \text{and}\quad
\pi^{(2)}_{k}= \pi^{(2)}_\alpha\vert_{\F_k^2}.
\end{equation}
At this point it is useful to remark that $$
\F^3_{K}=\{(k_\alpha,k_\beta ,k_\gamma )\in
({\bbT}^3)^2: k_\alpha+k_\beta +k_\gamma = K
\}\quad K \in {\bbT}^3
$$ and $$ \F^2_{k}=\{(k_\beta ,k_\gamma )\in ({\bbT}^3)^2: k_\beta
+k_\gamma = k  \},\quad k \in {\bbT}^3
$$ are six and three-dimensional manifolds isomorphic to
${({\bbT}^3)^2}$ and ${\bbT}^3,$ respectively.

\begin{lemma}
The mappings $\pi^{(3)}_{K}$ , $K\in \T^3$ and $\pi^{(2)}_{k}$,
$k\in \T^3$ are bijective from $\F_K^3\subset (\T^3)^3$	 and
$\F_k^2\subset (\T^3)^2$ onto $(\T^3)^2$ and $\T^3$ with the
inverse mappings given by
\begin{equation*}
(\pi^{(3)}_{K})^{-1}(q_\alpha,p_\alpha)= (l_{\alpha} K-p_\alpha,
l_{\beta} K+l_{\gamma\beta}p_\alpha+ q_\alpha, l_{\gamma} K+
l_{\beta\gamma}p_\alpha- q_\alpha)
\end{equation*}
and
\begin{equation*}
(\pi^{(2)}_{k})^{-1}(q_\alpha)=(l_{\gamma\beta}k+
q_\alpha,l_{\beta\gamma}k-q_\alpha).
\end{equation*}
\end{lemma}
\begin{proof} We obviously have that
\begin{equation*}
(l_{\alpha} K-p_\alpha)+ (l_{\beta} K+l_{\gamma\beta}p_\alpha+
q_\alpha)+ ( l_{\gamma} K+ l_{\beta\gamma}p_\alpha- q_\alpha)=K
\end{equation*}
and
\begin{equation*}
(l_{\gamma\beta}k+ q_\alpha)+(l_{\beta\gamma}k-q_\alpha)=k.
\end{equation*}
Therefore, the images of the mappings $(\pi^{(3)}_{K})^{-1}$ and
$(\pi^{(2)}_{k})^{-1}$ are the subsets of $\F^3_K$ and $\F^2_k,$
respectively.

Conversely, given
\begin{equation*}
(k_\alpha, k_\beta, k_\gamma)\in {\F}^3_K\subset (\T^3)^3 \quad
\text{and}\quad
 (k_\beta, k_\gamma)\in \F^2_k\subset (\T^3)^2
\end{equation*}
one computes that
\begin{align*}
(\pi^{(3)}_{K})^{-1}(q_\alpha,p_\alpha)= (k_\alpha, k_\beta,
k_\gamma) \quad \text{ and } \quad
(\pi^{(2)}_{k})^{-1}(q_\alpha)=(k_\beta, k_\gamma),
\end{align*}
where
\begin{align*}
& q_\alpha= l_{\beta\gamma} k_\beta-l_{\gamma\beta} k_\gamma \quad
\text{ and } \quad p_\alpha= l_\alpha
(k_\beta+k_\gamma)-(l_\beta+l_\gamma)k_\alpha  .
\end{align*}
\end{proof}

Let the
 operator $ H(K),$ $K \in {\T}^3$ act on the Hilbert space
 $L_2((\T^3)^2)$ as follows:
 $$
H(K)=H_0(K)-V_1-V_2- V_3.
 $$
The operators $H_0(K)$ and $V_\alpha$ in the coordinates
$(q_\alpha,p_\alpha)$
  are defined  by
$$(H_0(K)f)(q_\alpha,p_\alpha)=E_{\alpha\beta }
(K;q_\alpha,p_\alpha)f(q_\alpha,p_\alpha),\quad f\in L_2 ((
{\T}^3)^2), $$
 \be\no (V_\alpha
f)(q_\alpha,p_\alpha)=(2\pi)^{-\frac{3}{2}} \int\limits_{{\bbT}^3}
v_{\alpha}(q_\alpha-q'_\alpha)
f(q'_\alpha,p_\alpha)dq'_\alpha,\quad f\in L_2 (({\T}^3)^2), \ee
 where
$$ E_{\alpha\beta}(K;q_\alpha, p_\alpha)= \varepsilon _\alpha
(l_{\alpha}K-p_\alpha)+\varepsilon _\beta (l_{\beta} K
+l_{\gamma\beta} p_\alpha+q_\alpha) +\varepsilon _\gamma
(l_{\gamma} K+l_{\beta\gamma} p_\alpha-q_\alpha). $$

Let the
 operators $h_{\alpha}(k),\,\alpha=1,2,3,\,k \in {\T}^3$ acts on the Hilbert space
 $L_2(\T^3)$ as follows:
\begin{equation}\label{two} h_\alpha (k) =h_\alpha^{0}
(k)-v_{\alpha},
\end{equation} where
$$(h_\alpha^{0}
(k)f)(q_\alpha)=E_{k}^{(\alpha)}(q_\alpha)f(q_\alpha),\quad f \in
L_2({\bbT}^3), $$
\begin{equation}\no (v_\alpha
f)(q_\alpha)=(2\pi)^{-\frac{3}{2}} \int\limits_{{\bbT}^3}
v_{\alpha}(q_\alpha-q_\alpha') f(q_\alpha')d q_\alpha', \quad f \in L_2({\bbT}^3)
\end{equation}
and
\begin{equation}\label{E-alpha}
E_{k}^{(\alpha)}(q_\alpha)= \varepsilon _\beta
(l_{\gamma\beta}k+q_{\alpha}) +\varepsilon _\gamma
(l_{\beta\gamma}k-q_{\alpha}).
\end{equation}

Let us consider the	 unitary operators
 $$ U_K :L_2({ {\F}^3_K})
\longrightarrow L_2(({\T}^3)^2) ,
 \,\, U_K f=f \circ (\pi^{(3)}_{K})^{-1},
\,\, K \in {\bbT}^3, $$ and $$ u_{k}:L_2(\F^2_{k}) \rightarrow
L_2({\bbT}^3),	\,\, u_{k} g=g \circ(\pi^{(2)}_{k})^{-1},\,k\in
{\bbT}^3, $$ where $\pi^{(3)}_{K}$ and $\pi^{(2)}_{k}$ are defined
by \eqref{project}.

For	 the fiber operators $\widetilde H(K)$
resp. $\tilde h_\alpha(k)$ the following equalities
\begin{equation*}
H(K) = U_K \widetilde H(K)U_K^{-1},\quad h_\alpha (k) =u_{k}
\tilde h_\alpha (k)u_{k}^{-1},\quad \alpha=1,2,3
\end{equation*}
hold.

All of our further calculations will be carried out in the
"momentum representation" in a system of coordinates connected
with the fixed center of inertia of the system of three particles.
 We order 1,2,3 by the
conditions $1 \prec 2,$ $2 \prec 3$ and $3 \prec 1.$
 Sometimes instead of the
coordinates $(q_{\alpha},p_{\alpha})$ (if it does not lead to any
confusion we will write $(q,p)$ instead of $(q_\alpha,p_\alpha)$)
it is convenient to choose some pair of the three variables
$p_{\alpha}.$ The connection between the various coordinates is
given by the relations
\begin{equation}\label{coordinate}
p_1+p_2+p_3=0,\\ \pm q_\alpha=l_{\gamma\beta}p_\alpha + p_\beta,
\quad l_{\gamma\beta}=\frac{m_\beta}{m_\beta+m_\gamma},\quad
(\alpha\neq\beta, \beta\neq \gamma,\gamma\neq\alpha),
\end{equation} where the plus sign corresponds to the case $\beta
\prec \alpha,$ the minus sign corresponds to the case $\alpha
\prec \beta.$ Expressions for the variables $q_{\alpha}$ in terms
of $p_{\alpha}$ and $p_{\beta}$ can be written in the form
$q_{\alpha}=d_{\alpha\beta}p_{\alpha}+e_{\alpha\beta}p_{\beta}$
and explicit formulas for the coefficients $d_{\alpha\beta}$ and
$e_{\alpha\beta}$ are obtained by combining the latter equation
with \eqref{coordinate}.

\section{Statement of the main results}
For each $K\in\T^3$ the minimum and the maximum taken over $(q,p)$
of the function $E_{\alpha \beta}(K;q,p) $ are independent of ${\
\alpha },{\ \beta } =1,2,3.$ We set:
\begin{align*}
E _{\min }(K)\equiv\min_{q,p}E_{\alpha \beta }(K,q,p),\quad E_{max
}(K)\equiv\max_{q,p}E_{\alpha \beta }(K,q,p).
\end{align*}

The main results of the paper are given in the following theorems,
which will be proven in section 5, 7.

\begin{theorem}\label{exi} Assume Hypothesis \ref{eps} and \ref{hy}.
Then for any $\alpha=1,2,3$
and for all $ k \in U_\delta(0),$ $\delta>0$ sufficiently small,
 the operator $h_\alpha(k)$ has a finite number of eigenvalues
outside of the
essential spectrum	$\sigma _{ess} (h_{\alpha}(k))$.
\end{theorem}

\begin{theorem}\label{esss} Assume Hypothesis \ref{eps} and \ref{hy}. For the essential spectrum
${\sigma}_{ess} ( H(K))$ of $H(K)$ the following equality $$
\sigma_{ess}(H (K))=\cup^3_{\alpha=1}\cup _{p\in {\bbT}^3}\{\sigma
_d(h_{\alpha}
 ((l_\beta+l_\gamma)K+p))+\varepsilon _\alpha
(l_{\alpha} K-p)\} \cup [E_{\min}(K),E_{\max}(K)] $$ holds, where
$\sigma _d (h_{\alpha}(k))$ is the discrete spectrum of the
operator $h_{\alpha}(k),k\in \T^3$.
\end{theorem}

\begin{theorem}\label{finite} Assume Hypothesis \ref{eps} and \ref{hy}. The essential spectrum
${\sigma}_{ess} ( H(K))$ of $H(K)$ consists of the union of a
finite number of bounded closed intervals (segments).
\end{theorem}

\section{ Spectral properties of the
two-particle operator $h_\alpha(k)$}

In this section we study the spectral properties of the
two-particle discrete Schr\"{o}dinger operator $h_\alpha(k),\,\alpha=1,2,3,$
$k\in {\T^3}$ defined by \eqref{two}.

By the Weyl theorem the continuous spectrum
   $\sigma_{\text{cont}}(h_\alpha(k))$ of the
operator $h_\alpha(k),k \in \T^3$ coincides with the spectrum $
{\sigma}( h^0_\alpha (k) ) $ of $h^0_\alpha(k).$ More
specifically, $$
 \sigma_{\text{cont}}(h_\alpha(k))= [E^{(\alpha)}_{\min}(k)
,E^{(\alpha)}_{\max}(k)], $$
 where
\be\label{min} E^{(\alpha)}_{\min }(k)\equiv\min_{p\in
\T^3}E^{(\alpha)}_k(p),\quad
E^{(\alpha)}_{\max}(k)\equiv\max_{p\in {\bbT}^3}
E^{(\alpha)}_{k}(p) \ee
and $E^{(\alpha)}_{k}(p)$ is defined by \eqref{E-alpha}.

\begin{lemma}\label{hyp0} Assume Hypothesis \ref{eps}. Then	 the functions
 $\varepsilon_{\alpha}(p),\alpha=1,2,3$ defined on
 $\R^3$ are	 even,	 real-analytic
and	 the point $p=0$ is its unique	 non-degenerate minimum in
$\T^3$ .
\end{lemma}
\begin{proof}
The conditions $\mbox{(i)}$ and $\mbox{(ii)}$ of Hypothesis
\ref{eps} and the properties of Fourier transform implies that the
function $\varepsilon_{\alpha}(p)$ is even and real-analytic.

By	 $\mbox{(i)}$ of Hypothesis \ref{eps} the function
$\varepsilon_{\alpha}(p)$ is represented as
 \be \label{kinet1}
\varepsilon_{\alpha}(p) = \sum_{s\in \Z^3} \hat
\varepsilon_\alpha(s)e^{i(p,s)}= \sum_{s\in \Z^3} \hat
\varepsilon_\alpha(s)\cos (s^{(1)} p^{(1)} )\cos (s^{(2)} p^{(2)}
)\cos (s^{(3)} p^{(3)} ). \ee

From the representation \eqref{kinet1} we obtain  that for the
second-order partial derivatives of $\varepsilon_{\alpha}(p)$ at
the point $p=0$	 the equalities
\be \label{partial} \frac{\partial^2
\varepsilon_{\alpha}}{\partial p^{(i)}\partial p^{(j)}}(0)=0,\, i
\neq j,\quad \frac{\partial^2 \varepsilon_{\alpha}}{\partial
p^{(i)}\partial p^{(i)}}(0)=\frac{1}{m_{\alpha}}, \quad i,j=1,2,3,
\ee
 hold,
where the number $m_{\alpha}>0$ is defined in Remark \ref{mas}.
 Hence the Taylor series expansion of $\varepsilon_{\alpha}(p)$
 at the point $p=0$ gives us
\be \label{taylor}
\varepsilon_{\alpha}(p)=\varepsilon_{\alpha}(0)+\frac{p^2}{2m_{\alpha}}+
\tilde \varepsilon_{\alpha}(p),\quad \tilde
\varepsilon_{\alpha}(p)=O(|p|^4)\quad \mbox{as }\quad p\to 0. \ee
The equality \eqref{taylor} yields that the point $p=0$ is a
non-degenerated minimum of the function $\varepsilon_{\alpha}(p)$.

Therefore according to \eqref{kinet1} we get \be \label{kinet2}
\varepsilon_{\alpha}(p)-\varepsilon_{\alpha}(0)	 =- \sum_{s\in
\Z^3} \hat \varepsilon_{\alpha}(s)[1-\cos (s^{(1)} p^{(1)} ) \cos
(s^{(2)} p^{(2)} )\cos (s^{(3)} p^{(3)} )]. \ee The condition
$\mbox{(iii)}$ of Hypothesis \ref{eps} and \eqref{kinet2} implies
that $p=0$ is the unique non-degenerated minimum of the function
$\varepsilon_{\alpha}(p)$ in $\T^3$.
\end{proof}

\begin{lemma}
\label{minimum}
There exist an	analytical function
$p_ { \alpha } (k)$ defined	 on $ { \delta } $ - neighborhood
 $U_ {\delta } (  0 ) $ of the point $p=0$ such
  that for any $k { \in } U_ { \delta }(0)$	 the point $p_ { \alpha } (k) $ is an unique
  non-degenerate minimum  of the function
$E_k^{(\alpha)}(p)$.
\end{lemma}

\begin{proof}
Since the function $ {
\varepsilon } _ { \alpha } (p), { \alpha } =1,2,3$
 has a unique non degenerate  minimum at the point $p=0,$
 the gradient
  $ { \bigtriangledown } { \varepsilon } _ { \alpha } (
p ) $ is equal to  zero at the point $p=0,$ i.e. $$
 {\bigtriangledown }
  { \varepsilon } _ { \alpha } (p) { \vert }_{p= 0} =
  ( \frac
  { \partial  \varepsilon _ \alpha (p) }
   {\partial  p^{(1)}},
\frac { \partial  \varepsilon _ { \alpha } (p) }
 {{ \partial }
p^{(2)}},
 \frac {{ \partial
\varepsilon } _ { \alpha } (p)} { \partial p^{(3)} })
 { \vert }_{p= 0}=0.
$$

Therefore by  \eqref{partial} the matrix
 $$ B_ { \alpha }
 (p )|_{p=0}=
\big ( \frac {{ \partial } ^2 { \varepsilon } _ { \alpha } ( { 0 }
 ) } {{ \partial } p^ { (i) } { \partial } p^ { (j) }}
\big )
 { \vert } _{_ { i, j=1,2,3 }}=
 m^{-1}_\alpha I_3
 ,\,
\alpha=1,2,3,
 $$
is positive, where $I_3$ is the	 $3\times 3$ unit matrix.

From here it follows that $ {\bigtriangledown } { E } _
0^{(\alpha)}(0) = 0 $
 and
the matrix $ B (0 )
 =\big ( \frac {{ \partial } ^2
  { E
}_0^{(\alpha)} (0 )}
 {{\partial } p^{(i)} { \partial } p^{(j)}}
 \big ) { \vert } _ { i, j=1 }
^3 = (m^{-1}_\beta+m^{-1}_\gamma)I_3 $ is positive definite.
 Now we will apply the implicit function  theorem
	  to the equation
	 $ {\bigtriangledown } {E } _k^{(\alpha)}  ( p)=0
,\,\,k,p\in {\T}^3.$
		 Then
there exists a vector  function $p_\alpha (k)$	defined and
analytic in some $ { \delta } $-neighborhood  $U_ { \delta } (0 )
$ of the point $k=0$, and for any $k { \in } U_ { \delta } ( 0 ) $
the equality
$ { \bigtriangledown } E
_k^{(\alpha)}  ( p_\alpha(k))
 = 0 $
holds.

Denote by $B
(k) $ the matrix of the second order partial derivatives of the
function $ E_k^{(\alpha)} ( p) $
 at the point
$ p_{\alpha}(k)$.

Since the matrix $B(0)$ is positive and	 $B (k)$ is
continuous in $U_ { \delta } ( 0 )$, we conclude that for any $k
\in U_ { \delta } ( 0 )$
 the matrix $B (k)$ is positive definite. Thus
$ p_{\alpha}(k),\,k \in U_ { \delta } ( 0 )$ is the unique
non-degenerated minimum point
 of $E_k^{(\alpha)} (p).$
\end{proof}


Let $\C$ be the complex plane. Denote by $r^0_\alpha(k,z)$ the
resolvent of the operator $h^0_\alpha(k),\,\alpha=1,2,3$.
 For any $k \in \T^3$ denote by
$\Delta_{\alpha}(k,z)$ the Fredholm determinant of the	operator
\begin{equation}\label{fred}
I-v_{\alpha}r^0_\alpha(k,z),\,\quad z \in \C\setminus
[E^{(\alpha)}_{\min}(k) ,E^{(\alpha)}_{\max}(k)],
\end{equation}
where $I$ is the identity operator on $L_2(\T^3).$

\begin{lemma}\label{Det}
 For any $k\in {\T}^3$ the number
 $z\in \C \setminus [E^{(\alpha)}_{\min}(k)
,E^{(\alpha)}_{\max}(k)]$ is an eigenvalue of the operator
$h_\alpha(k),\,\alpha=1,2,3$ if and only if	 $$\Delta_\alpha(k,z)=0.$$
\end{lemma}
\begin{proof}
By	the Birman-Schwinger principle the number $z\in \C \setminus
[E^{(\alpha)}_{\min}(k) ,E^{(\alpha)}_{\max}(k)]$
 is an
eigenvalue of the operator $h_\alpha(k),\,\alpha=1,2,3,\,k \in \T^3$ if and only if
  the equation
\begin{equation}\label{equ}
g=v_{\alpha}r^0_\alpha(k,z)g
\end{equation}
has a nontrivial solution $\hat g \in L_2(\T^3)$.

By Fredholm's theorem the equation \eqref{equ} has nontrivial
solutions
 if and only if
$$\Delta_\alpha(k,z)=0.$$
\end{proof}

Now we obtain a generalization	of the Birman-Schwinger principle for the
discrete two-particle Schr\"odinger
operators and hence we prove Theorem  \ref{exi}.

 Let $N(k,z)$ denote the number
of eigenvalues of the operator $h_\alpha(k),\,k\in {\bbT}^3$ below
$ z\leq E^{(\alpha)}_{\text{min}}(k)$.
For any bounded self-adjoint operator $A$ acting in the Hilbert
space ${\cH}$ not having any essential spectrum on the right of
the point $z$ we denote by	${\cH}_A(z)$ the subspace such that
$(Af,f) > z(f,f)$ for any $f \in {\cH}_A(z)$ and set
$n(z,A)=\sup_{\cH_A(z)}\dim{\cH}_A(z)$.

For any $k\in U_{\delta}(0)$ and $z\leq
E^{(\alpha)}_{\text{min}}(k)$ we define the integral operator
$G_\alpha(k,z)$ with the kernel
\begin{equation}\no\label{yadroG}
G_\alpha(k,z;p,q)=(2\pi)^{-\frac{3}{2}}
\frac{v_{\alpha}(p-q)}{(E^{(\alpha)}_k(p)-z)^\frac{1}{2}
(E^{(\alpha)}_k(q)-z)^\frac{1}{2} }
\end{equation}

The following theorem is a realization of the  well known
Birman-Schwinger principle for the two-particle Schr\"{o}dinger
operators on  lattice $\Z^3$ .

\begin{theorem}\label{l.b-s} The operator $G_\alpha(k,z),\,k\in U_\delta(0)$
 acts in $ L_2({\T}^3)$, is
positive, belongs to the Hilbert-Schmidt class $\Sigma_2$ and is
continuous in $z$ up to $z=E^{(\alpha)}_{\text{min}}(k).$ In
addition the equality
\begin{equation}\no \label{b-s=}
N(k,z) =n(1,G_\alpha(k,z)),\,\,z \leq E^{(\alpha)}_{\text{min}}(k)
\end{equation} holds.
\end{theorem}
\begin{proof}
We note that for all $z< E^{(\alpha)}_{\text{min}}(k)$ the
equality $G_\alpha(k,z)=(r^{0}_\alpha(k,z))^{\frac{1}{2}} v_\alpha
(r^{0}_\alpha(k,z))^{\frac{1}{2}} $ holds.

The quantity $N(k,z)$ coincides with $n(1,G_\alpha(k,z))$ by the
Birman-Schwinger principle, i.e,
\begin{equation}\no \label{tenglik}
N(k,z)=n(1,G_\alpha(k,z)),\,z <E^{(\alpha)}_{\text{min}}(k).
\end{equation}

  By
Hypothesis \ref{hy}	 the function $v_\alpha(p)$ is continuous
 on $\T^3$. Since for any  $k\in
U_{\delta}(0)$ the function $E^{(\alpha)}_k(q)$ has a unique
non-degenerate minimum at the point $p=p_{\alpha}(k)$ the kernel
$G_\alpha(k,E^{(\alpha)}_{\text{min}}(k);p,q)$ is  square
integrable on $(\T^{3})^{2},$ that is, the operator
$G_\alpha(k,E^{(\alpha)}_{\text{min}}(k))$ belongs to the
Hilbert-Schmidt class $\Sigma_{2}.$

Then the dominated convergence theorem implies that
$G_\alpha(k,z)$ is continuous from the left up to
$E^{(\alpha)}_{\text{min}}(k)$.

 Let us show  $N(k,E^{(\alpha)}_{\text{min}}(k))
=n(1,G_\alpha(k,E^{(\alpha)}_{\text{min}}(k))).$
 Since $G_\alpha(k,E^{(\alpha)}_{\text{min}}(k))$ is a
compact operator the number
$n(1-\gamma,G_\alpha(k,E^{(\alpha)}_{\text{min}}(k)))$ is finite
for any $\gamma < 1$.

Then according to the Weyl inequality
$n(\lambda_1+\lambda_2,A_1+A_2)\leq
n(\lambda_1,A_1)+n(\lambda_2,A_2)$ for all
$z<E^{(\alpha)}_{\text{min}}(k)$ and $\gamma \in (0,1)$ we have
$$
N(k,z)=n(1,G_\alpha(k,z))\leq
n(1-\gamma,G_\alpha(k,z))+n(\gamma,G_\alpha(k,z)-G_\alpha(k,E^{(\alpha)}_{\text{min}}(k))).
$$
Since $G_\alpha(k,z)$ is continuous from the left up to
$E^{(\alpha)}_{\text{min}}(k)$ we obtain
$$
\lim_{z\to E^{(\alpha)}_{\text{min}}(k)-0} N(k,z)=
N(k,E^{(\alpha)}_{\text{min}}(k))\leq
n(1-\gamma,G_\alpha(k,E^{(\alpha)}_{\text{min}}(k)))\,\, \mbox{for
all}\,\, \gamma \in (0,1)
$$
and so $$N(k,E^{(\alpha)}_{min}(k))\leq \lim _{\gamma\to
0}n(1-\gamma,G_\alpha(k,E^{(\alpha)}_{min}(k)))<\infty.$$ Since
$N(k,E^{(\alpha)}_{\text{min}}(k))$ is finite we have
$N(k,E^{(\alpha)}_{\text{min}}(k)-\gamma)=N(k,E^{(\alpha)}_{\text{min}}(k))$
for all small enough $\gamma \in (0,1)$. Therefore using  the
continuity of $N(k,z)$ from the left we derive the equality
$$n(1,G_\alpha(k,E^{(\alpha)}_{\text{min}}(k)))=
\lim _{\gamma\to 0}n(1,G_\alpha(k,E^{(\alpha)}_{\text{min}}(k)
-\gamma))$$
$$=
\lim _{\gamma\to 0}N(k,E^{(\alpha)}_{\text{min}}(k)-\gamma)=
N(k,E^{(\alpha)}_{\text{min}}(k)).$$
\end{proof}
{\bf  Proof of Theorem \ref{exi}.}
Since $v_\alpha$ is positive we conclude that $h_\alpha(k)$ has no
eigenvalue lying on	 the  r.h.s	 of $\sigma_{cont}(h_\alpha(k)).$
Then the finiteness of the discrete spectrum
$\sigma_{d}(h_\alpha(k))$ of $h_\alpha(k)$ follows from the
compactness of $G_\alpha(k,E^{(\alpha)}_{\text{min}}(k))$ and
Lemma \ref{l.b-s}.
\begin{flushright} $\square$ \end{flushright}

\section{ Spectrum of  "channel operator"}
In this section we introduce a "channel operator" and
prove that its spectrum consists of only a	finitely many
segments.

The "channel operator" $H_{\alpha} ( K), K {\ \in } {\T}^3$ acts
in the Hilbert space $L_2 (({\T}^3 )^2) $ as
\begin{equation}\no
 H_{\alpha}(K)=H_0(K)- V_{\alpha}.
\end{equation}

The decomposition of the space $L_2 (( {\T}^3 ) ^2) $ into the
direct integral
 $$L_2 (( {\T}^3 ) ^2) = \int\limits_{p\in {\T}^3}
\oplus L_2({\T}^3) d p $$ yields for the operator $H_{\alpha}(K)$
the decomposition into the direct integral
 $$ H_{\alpha}(K) =	 \int\limits_{p\in {\T}^3}
 \oplus H_{\alpha}(K,p) dp.$$
 The fiber operator $H_{\alpha}(K,p)$ acts in the
Hilbert space $L_2({\bbT}^3)$ and has the form
\begin{equation}\label{chan.fiber}
 H_{\alpha}(K,p) =h_{\alpha}
((l_\beta+l_\gamma)K+p)+\varepsilon_{\alpha} (l_{\alpha}K-p)
I,
\end{equation} where  $h_{\alpha}(k)$ is the
two-particle operator defined by \eqref{two}.

Set
\begin{align*}
&\Delta_{\alpha}(K,p,z)=\Delta_{\alpha}\Big
((l_{\beta}+l_{\gamma})K+p,
z-\varepsilon_{\alpha}(l_{\alpha}K-p) \Big ),\no\\
&E^{(\alpha)}_{\min}(K,p)=E^{(\alpha)}_{\min} \Big
((l_{\beta}+l_{\gamma})K+p\Big )+
\varepsilon_{\alpha}(l_{\alpha}K-p ),\no\\
&E^{(\alpha)}_{\max}(K,p)=E^{(\alpha)}_{\max} \Big
((l_{\beta}+l_{\gamma})K+p \Big)+
\varepsilon_{\alpha}(l_{\alpha}K-p),
\end{align*}
where $\Delta_{\alpha}(k,z)$ is the Fredholm determinant of the
operator $I-v_{\alpha}r^0_{\alpha}(k,z)$ and
$E^{(\alpha)}_{\min}(k)$ and $E^{(\alpha)}_{\max}(k)$ are defined in
\eqref{min}.

\begin{lemma}\label{Det1}
 For any $p\in {\T}^3$ the number
 $z\in \C \setminus [
E^{(\alpha)}_{\min}(K,p),E^{(\alpha)}_{\max}(K,p)]$ is an
eigenvalue of the operator $H_\alpha(K,p)$ if and only if
$$\Delta_{\alpha}(K,p,z)=0,\,\alpha=1,2,3.$$
\end{lemma}
The proof of Lemma \label{Det1} similarly to that Lemma\ref{Det}.

The representation of the operator $H_\alpha (K,p)$	 implies the
equality
\begin{align*}\label{stucture}
 &\sigma (H_\alpha (K,p))
 = \{\sigma _d \big (h_{\alpha}((l_\beta+l_\gamma)K+p)\big )+\varepsilon _\alpha
(l_{\alpha}K-p)\}\\
 &\cup [E_{\min
}^{(\alpha)} (K,p ),E_{\max}^{(\alpha )}(K,p)],
\end{align*}
 where $\sigma _d(h_{\alpha}(k))$
 is the discrete spectrum of the operator
 $h_{\alpha}(k).$
Set
$$
\sigma _{two}(H_\alpha (K))=
 \bigcup _{p\in {\T}^3}\{\sigma _d\big (h_\alpha
((l_\beta+l_\gamma)K+p)\big )+\varepsilon _\alpha
(l_{\alpha}K-p)\},\,\alpha=1,2,3 .
$$

The theorem (see,e.g.,\cite{RidSimIV}) on the spectrum of decomposable
operators and above obtained structure for the spectrum of
$H_\alpha (K,p)$ gives
 \begin{lemma}\label{spec} The
equality holds
\begin{align*}
&\sigma (H_\alpha (K))=\sigma _{two}(H_\alpha (K)) \cup
[E_{\min}(K),E_{\max}(K)].
\end{align*}
\end{lemma}

\begin{theorem} \label{set1} The set
\be\no \label{set}
\hat \sigma_\alpha =\overline{\sigma_{two}(H_\alpha(K)) \setminus
[E_{min}(K),E_{max}(K)]}
\ee
 consists of an union of a finite number of segments.
\end{theorem}

\begin{proof}
Let the set	 $\hat \sigma_\alpha$ be represented as the union of a finite or infinite
number of disjoint segments $S_{\omega}$  represented in the form
$$ \hat \sigma_\alpha=
\bigcup_{\omega \in W} S_\omega,$$ where $W$ is a subset of the
real axis.

Denote by $d_\omega$ the  distance between of the segments
 $S_\omega$ and $[E_{min}(K),E_{max}(K)]$.

\begin{lemma}\label{exis} Let  $d_\omega >0$  for some $\omega \in W.$	Then  for any  $p \in {\T}^3$
the operator $H_\alpha(K,p)$ \eqref{chan.fiber} has	 eigenvalues in
$S_\omega.$
\end{lemma}
\begin{proof}
 Let $d_{\omega}>0$ for some $\omega \in W.$
 Denote by $G_\omega$ the set of all $p \in {\T}^3$ such that
the operator $H_\alpha(K,p)$ has an eigenvalue lying in
$S_\omega.$ We show that $G_\omega=\T^3.$
Let $p_0 \in G_\omega$ then by Lemma   \ref{spec} and \ref{Det1}
there is $z_0 \in S_\omega$ such that
 $\Delta_\alpha (K,p_0,z_0)=0.$

For the following considerations in this proof we shall consider in $\T^3\equiv (-\pi,\pi]^3$
as equipped with the topology of the corresponding 3-dimensional torus, and
vice versa.
For any $p \in {\T}^3$ the function	 $\Delta_\alpha(K,p,z)$
   is analytic by $z$ in some region  containing
$S_\omega$ and nonzero. Therefore there is natural number $n$ such
that the inequality $\frac{\partial ^n}{\partial  z
^n}\Delta_\alpha (K,p_0,z_0) \neq 0$ holds.
 By the implicit function theorem  there exist a neighborhood $U(p_0)$ of
$p_0$ and a continuous function	 $z(p) \in S_\omega$ defined on
$U(p_0)$
 such that the identity
$\Delta_\alpha(K,p,z(p))\equiv 0$ is valid. According to Lemma
\ref{Det1} the number $z(p) \in S_\omega$ is an eigenvalue of
$H_\alpha(K,p)$ for $p \in U(p_0)\subset G_{\omega}.$ This means
that the set $G_\omega$ is open.

Now we prove that $G_\omega$ is a closed set.

Indeed, let	 $\{p_n\} \subset G_\omega$ converge to $p_0 \in
{\T}^3$ and let $\{z(p_n)\} \subset S_\omega$ be eigenvalues of
$H_\alpha(K,p_n).$	Let	 $z_0 \in S_{\omega}$ be a limit point of
$\{z(p_n)\}$.

 The function $\Delta_\alpha(K,p,z)$ is
continuous in $ (p,z)\in \T^3\times
 S_\omega$.
 Therefore $$
0\equiv\lim_{n \to \infty}
\Delta_\alpha(K,p_n,z(p_n))=\Delta_\alpha(K,p_0,z_0)
$$
and hence $p_0 \in G_\omega$. So the set $G_\omega$ is closed.
Since $G_\omega$ is an open and closed set we have
$G_\omega={\T}^3.$
\end{proof}
Now we prove that the set  $W$ is finite. Assume in fact, ad absurdum  that the set $W$
is	infinite, i.e., for infinitely many elements $\omega \in W$ one has
 $d_\omega >0$. Then by Lemma \ref{exis} for any $p \in
{\T}^3$ the operator $H_\alpha(K,p)$
 has an eigenvalue in $S_\omega,\, \omega \in W$.

Therefore by  the equality $\sigma_d(H_\alpha (K,p))=\sigma_d\big ( h_\alpha
((l_\beta+l_\gamma)K+p)\big )+\varepsilon _\alpha (l_{\alpha}K-p) $
 for any $p \in {\T}^3$ the set $ \sigma_d (h_{\alpha }
(p))$ is infinite. By Theorem \ref{exi} for all $k\in U_\delta(0)$
the operator
 $h_{\alpha } (k)$ has
finitely many eigenvalues. This is in contradiction with
our assumption
\end{proof}

\section{Essential spectrum of the discrete three-particle Schr\"odinger
operator $H(K)$}

In this section
we prove  Theorem \ref{esss}
and \ref{finite}
using the
 Faddeev type system of integral equations.

{\it \bf Proof of Theorem \ref{esss}.} Set
 $$
\Sigma(K)= \cup_{\alpha=1}^3 \sigma _{two}(H_\alpha (K)) \cup
[E_{min}(K),E_{max}(K)].
$$

 We first  show that $\Sigma(K) \subset \sigma_{ess}(H(K)).$
 Let $z_0 \in \Sigma(K)$ be arbitrary point.
We construct an orthogonal	sequence of functions $\{f_n\}_{1}^{\infty}$ such that
$$ ||(H(K)-z_0)f_n|| \to 0 \quad \mbox{as}\quad	 n \to \infty. $$

 For any $z_0\in [E_{min}(K), E_{max}(K)]$ there exists $(q_{0},p_0)\in
(\T^3)^2$ such that $z_0=E_{\alpha\beta}(K;q_{0},p_{0}).$

For any	 $n\in \N$	and $p,q\in \T^3$ we introduce the notations:
\begin{equation}\label{nigh}
U_n:=  U_n(p)=\{x\in \T^3: \frac {1}{n+1} < |x-p| < \frac {1}{n}
\}\,\, \mbox{and}\,\, W_n:= W_n \big ( q,p\big )=U_n(q)\times
U_n(p).
\end{equation}

Let
$$
 f_n(q,p)= \left \{
\begin{array}  {ll}
\frac{1}{\sqrt{\mu (W_n)}},\qquad	 \mbox{if}	\quad  (q,p) \in
W_n(q_0,p_0) \\ 0,	   \qquad \mbox{if}	 \quad (q,p) \notin
W_n(q_0,p_0) ,
\end{array}
\right. $$ where  $ \mu (W_n)\,\,\mbox{ is the Lebesgue measure of } \,\, W_n\big(q_0,p_0 \big).$

 For any  $n \not=m$ the equality	 $W_n \cap W_m= \emptyset$	holds, hence
 $\{f_n\}_{1}^{\infty}$ is an orthogonal system.

 The norm $||(H(K)-z_0)f_n||$ is estimated by
\begin{equation}\label{est}
 ||(H(K)-z_0)f_n||^2
 \leq 2\left( ||(H_0(K)-z_0)f_n||^2+
\sum_{\alpha=1}^3||V_\alpha f_n||^2
 \right).
\end{equation}

 We shall prove	 that each item on the r.h.s. of
\eqref{est} tends to zero as $n \to \infty$.

Since  $E_{\alpha\beta}(K;q,p)$ is continuous and
$\sup\limits_{(q,p)\in W_n}|p-q|\to 0$ as $n \to \infty$, we have
\begin{equation}\no
||(H_0(K)-z_0)f_n||^2
 \leq \sup\limits_{(q,p)\in W_n}|E_{\alpha\beta}(K;q,p)-z_0|^2
 \to 0 \quad
\mbox{as}\quad n \to \infty.
\end{equation}

By	the Schwarz inequality	and the absolutely continuity of the
Lebesgue integral we have
 $$
 \int\limits_{(\T^3)^2} |(V_\alpha f_n)(q,p)|^2dpdq \leq
  (2\pi)^{-3} \int\limits_{\T^3} \int\limits_{U_n}|v_\alpha(t-p)|^2dpdt
   \to 0 \quad \mbox {as} \quad n \to \infty.
$$

By Weyl's criterion $z_0 \in \sigma_{ess}(H(K)).$

  Since $z_0 \in
[E_{min}(K),E_{max}(K)]$ is arbitrary, we have
$[E_{min}(K),E_{max}(K)] \subset \sigma_{ess}(H(K)).$

Let us show that $\sigma_{two}(H_{\alpha}(K)) \subset
\sigma_{ess}(H(K)).$ Let $z_0 \in \sigma_{two}(H_{\alpha}(K))$.
  By Lemma \ref{Det1} and Lemma \ref{spec} there exists
$p_0 \in {\T}^3$ such that $\Delta_\alpha(K,p_0,z_0)=0$.

 By the	 implicit function theorem	there
are neighborhoods $U(p_0)\subset \T^3 $ and $V(z_0)\subset \R^1$
of $p_0$ resp. $z_0$  and continuous function
$z:U(p_0)\to V(z_0),$ satisfying the condition $ \Delta
(K,p,z(p))=0 \, \mbox {and} \, z(p) \in V(z_0). $ It is clear
$z(p)$ is an eigenvalue of the operator $H_\alpha(K,p)$ for $p \in
U(p_0).$

Let for $p \in
U(p_0)$ the function $ f_p(q)$	be an eigenfunction of $H_\alpha(K,p)$
corresponding to the eigenvalue $z(p)\in V(z_0)$, that is,
 \begin{equation}\no \label{ei.func}
H_{\alpha}(K,p) f_p(q)=z(p) f_p(q),\quad p \in	U(p_0).
\end{equation}

Let $\chi_{_{U'_n}}(p)$ be the characteristic function of the set
$U'_n(p_0)=U(p_0)\cap U_n(p_0),$ where $U_n(p_0)$ is defined by
\eqref{nigh}.

Define the function $f(q,p)$ on $({\T}^3)^2$ by
$$
 f(q,p)=\left\{
\begin{array}  {ll}
f_p(q),\,\, p\in U(p_0),\,\, q\in {\T}^3 \\ 0, \,\, p\in {\T}^3
\setminus U(p_0),\,\, q\in {\T}^3
\end{array}
\right. $$

It is obvious that $f\in L_2(({\T}^3)^2).$ Let $f_n(q,p) = C_n
\chi_{_{U'_n}}(p)f(q,p),$ where
$C_n=(\int\limits_{\T^3}|\chi_{_{U'_n}}(p)f(q,p)|^2dpdq)^{-\frac{1}{2}}>0$
for a sufficiently large $n \in \N$. Then $||f_n||=1.$

Since
\begin{equation}\label{ab}
 ||(H-z_0)f_{n}|| \leq ||(H_\alpha(K)-z_0)f_n|| +
||V_{\beta}f_n||+||V_{\gamma}f_n||,
\end{equation}
we will show that each term on the r.h.s. of \eqref{ab} tends to
zero as $n \to \infty.$ Since
 $$
 ||(H_\alpha(K)-z_0)f_n||^2
 \leq
 sup_{p\in U'_n(p_0)}|z(p)-z_0|^2 $$
and $z(p)$ is continuous in
$U(p_0),$ we obtain that $||(H_\alpha(K)-z_0)f_n||\to 0$ as
$n \to \infty.$

Using the  Schwarz inequality and
applying the absolutely continuity of the Lebesgue integral
 we
have $$ ||V_\beta f_n||^2\leq
(2\pi)^{-3}\int\limits_{\T^3}\int\limits_{U_n}|v_\beta (p-p')|^2dp'dp
 \to 0\quad \mbox {as} \quad n
\to \infty. $$

Thus $z_0 \in \sigma_{ess}(H(K)) $. Since  $z_0\in \sigma_{two}(H_{\alpha}(K))$
is arbitrary, we have $\sigma_{two}(H_{\alpha}(K))
\subset \sigma_{ess}(H).$

 So we proved the inclusion $\Sigma(K)	\subset
\sigma_{ess}(H(K))$.

Now we prove the following inclusion $	\sigma_{ess}(H(K)) \subset
\Sigma(K) $.

Let $R_\alpha(K,z)$ resp. $R_0(K,z)$ be the resolvents of the
operators $H_\alpha(K)$ resp. $H_0(K).$

Denote by $v_\alpha^{\frac{1}{2}}$ the positive square root of the
integral operator $v_\alpha$ and by $v_\alpha^{\frac{1}{2}}(p-p')$
the kernel of the integral operator $v_\alpha^{\frac{1}{2}}$.

Let us consider the operator acting in $L_2(({\T}^3)^2)$ as
\begin{align}\no \label{WV}
 &(V^{\frac{1}{2}}_\alpha f)
(q_\alpha,p_\alpha)=(v^{\frac{1}{2}}_\alpha \otimes I  ) f
(q_\alpha,p_\alpha),
\end{align}
 where $\otimes$ denotes the tensor product of
operators.

Let $W_\alpha(K,z),\alpha=1,2,3$ be the operators on
$L_2(({\T}^3)^2)$ defined as
 \begin{equation}\no \label{Wz}
W_\alpha(K,z)={\bf I}+V_\alpha^{\frac{1}{2}}
R_\alpha(K,z)V_\alpha^{\frac{1}{2}},
 \end{equation}
where ${\bf I}$ is the identity operator on $L_2((\T^3)^2).$
 One can check that
 \begin{equation}\no
W_\alpha(K,z)=({\bf I}-V^{\frac{1}{2}}_\alpha
R_0(K,z)V^{\frac{1}{2}}_\alpha)^{-1}.
 \end{equation}

Denote by $L^{(3)}_2(({\T}^3)^2)$ the space of vector
functions $w$, with components $w_{\alpha}\in L_2(({\T}^3)^2),
{\alpha}=1,2,3.$

Let	 $ T(K,z),\, z \in {\C} \setminus {\Sigma(K)}$
be the operator	 on $L^{(3)}_2(({\T}^3)^2)$ with  the entries
\begin{equation}\no \label{TKZ}
\left\{ \begin{array}{ll}
 T_{ \alpha\alpha } ( K, z) = 0, \\
 T_{ \alpha\beta } ( K, z) =
W_{\alpha}(K,z)V_\alpha^{\frac{1}{2}}
R_0(K,z)V_\beta^{\frac{1}{2}}, \,{\ \alpha } {\neq} {\ \beta
},{\alpha},{\beta}=1,2,3.
\end{array}\right.
\end{equation}

\begin{lemma}\label{tz} For any $z \in \C\setminus {\Sigma(K)}$
the operator $T(K,z)$ is an Hilbert-Schmidt operator.
\end{lemma}
\begin{proof}
Recall that the kernel function $v_\alpha^{\frac{1}{2}}(p)$
 of $V_\alpha^{\frac{1}{2}}$
 is	 the inverse Fourier
transform of the function $\hat v_{\beta\gamma}^{\frac{1}{2}}(s)$
and
 belongs to $L_2(\T^3).$
 Then
one can check that	for any $z \in \C\setminus {\Sigma(K)}$
 the operator $V_\alpha^{\frac{1}{2}}
R_0(K,z)V_\beta^{\frac{1}{2}},\,\alpha\neq \beta $ belongs to the
Hilbert-Schmidt class $\Sigma_2$. Since for any $z \in \C\setminus
{\Sigma(K)}$ the operator $W_\alpha(K,z)$ is bounded, the operator
$T_{\alpha\beta}(K,z)$ also belongs to $\Sigma_2.$
\end{proof}

Denote by $R(K,z)=(H(K)-z{\bf I})^{-1}$ the resolvent of the
operator $H(K).$
  We consider the  well known	resolvent equation
\begin{equation}\label{Rez}
R(K,z)=R_0(K,z)+R_0(K,z)\sum_{\alpha=1}^{3}V_{\alpha}R(K,z).
\end{equation}

 Multiplying  \eqref{Rez} from the left side to
 $V^{\frac{1}{2}}_{\alpha}$ and setting
  $\cR_{\alpha}(K,z)\equiv V^{\frac{1}{2}}_{\alpha}R(K,z)$
 we get the equation
\begin{equation}\no 
\cR_{\alpha}(K,z)=V^{\frac{1}{2}}_{\alpha}R_0(K,z)+V^{\frac{1}{2}}_{\alpha}R_0(K,z)\sum_{\beta=1}^{3}V^{\frac{1}{2}}_{\beta}
\cR_{\alpha}(K,z),
\end{equation}
i.e, the following system of three equations
\begin{equation}\label{Rez1}
({\bf I}-V^{\frac{1}{2}}_{\alpha}R_0(K,z)V^{\frac{1}{2}}_{\alpha})\cR_{\alpha}(K,z)=
V^{\frac{1}{2}}_{\alpha}R_0(K,z)+\sum_{\beta=1,\,  \beta \neq
\alpha}^{3} V^{\frac{1}{2}}_{\alpha}R_0(K,z)
V^{\frac{1}{2}}_{\beta} \cR_{\alpha}(K,z).
\end{equation}

 Multiplying the equality \eqref{Rez1} from the left by
 the operator
 $$
W_\alpha(K,z)=({\bf I}-V^{\frac{1}{2}}_\alpha
R_0(K,z)V^{\frac{1}{2}}_\alpha)^{-1}$$	we get the Faddeev type
equation
\begin{equation}\label{Rez2}
{\bf R}(K,z)={\bf R}_0(K,z)+T(K,z){\bf R}(K,z),
\end{equation}
where  ${\bf R}(K,z)=(\cR_{1}(K,z),\cR_{2}(K,z),\cR_{3}(K,z))$ and \\
${\bf R}_0(K,z)=(W_1(K,z) V^{\frac{1}{2}}_{1}R_0(K,z),W_2(K,z)
V^{\frac{1}{2}}_{2}R_0(K,z),\,W_3(K,z)
V^{\frac{1}{2}}_{3}R_0(K,z))$ are  vector operators.

 From \eqref{Rez} we have the following representation for the
 resolvent
\begin{equation} \label{Rez4}
R(K,z)=R_0(K,z)+R_0(K,z)\sum_{\alpha=1}^{3}V^{\frac{1}{2}}_{\alpha}\cR_\alpha(K,z).
\end{equation}

 Let ${\cI}$ be the	 identity operator in $L^{(3)}_2((\T^3)^2).$
 The operator $T(K,z)$ is a
compact operator-valued function on $ \C\setminus {\Sigma(K)}$ and
${\cI }-T(K,z)$ is invertible if $z$ is real and either very
negative or very positive.The analytic Fredholm theorem (see, e.g.,
Theorem $VI.14$ in \cite{RidSimIV}) implies that there is a
discrete set $S \subset \C\setminus {\Sigma(K)} $ so that $({\cI
}-T(K,z))^{-1}$ exists and is analytic in $\C\setminus (
\Sigma(K) \cap S)$ and meramorphic in $\C\setminus {\Sigma(K)}$
with finite rank residues.
Thus $({\cI}-T(K,z))^{-1} {\bf
R}_0(K,z)\equiv F(K,z)$
is analytic in $\C\setminus ( \Sigma(K) \cup S)$ with finite rank
residues at the points of $S$.

Let $z \notin S$, $\Img z\neq 0$, then by \eqref{Rez2} , \eqref{Rez4}
we have $F(K,z)={\bf R}(K,z).$
In particular,
$$
R(K,z)(H(K)-z{\bf I})=(R_0(K,z)+R_0(K,z)\sum_{\alpha=1}^{3}V^{\frac{1}{2}}_{\alpha}\cR_\alpha(K,z))
(H(K)-z{\bf I})={\it \bf I}.
$$

 By analytic
continuation, this holds for any $z \notin \Sigma(K) \cup S. $ We
conclude that, for any such $z,$ $H(K)-z{\bf I}$ has a bounded
inverse. Thus  $\sigma(H(K)) \setminus	\Sigma(K) $ consists of
 isolated points and  only the	frontier points of $\Sigma(K)$
are	 possible their limit points. Finally, since $R(K,z)$ has
finite rank residues at any point $z \in S$, we conclude that
$\sigma(H(K)) \setminus \Sigma(K) $ belongs to the discrete
spectrum $\sigma_{d}(H(K))$ of $H(K).$
\begin{flushright}
$\square$
\end{flushright}
{\it {\bf Proof of Theorem \ref{finite}}.}
Theorem \ref{finite} follows immediately from
 Theorem \ref{set1} and
Theorem \ref{esss}	.
\begin{flushright}
$\square$
\end{flushright}

{\bf Acknowledgement} The authors grateful to Prof. R.A.Minlos,
Prof.K.A.Makarov and Prof. A.K.Motovilov for useful discussions.

This work was supported by the DFG 436 USB 113/3 and DFG 436 USB
113/4 projects and the Fundamental Science Foundation of
Uzbekistan. The last two named authors gratefully acknowledge the
hospitality of the Institute of Applied Mathematics and of the
IZKS of the University Bonn.

 {\small


\begin{thebibliography}{99}


\bibitem{AlbLakzM}	{\sc S. Albeverio, S.N.Lakaev,Z.I.Muminov}:Schr\"{o}dinger
operators on lattices. The Efimov effect and discrete spectrum
asymptotics (to appear in Annals Inst.H.Poincare).

\bibitem{Enss}	{\sc V. Enss}:A Note on Hunziker's Theorem. Comm. Math. Phys. {\bf 52} (1977),
 233-238.

\bibitem{F}	 {\sc L. D. Faddeev}: Mathematical aspects of the three--body
problem in quantum mechanics. Israel Program for Scientific
Translations, Jerusalem, 1965.

\bibitem{FM}  {\sc L. D. Faddeev} and {\sc S. P. Merkuriev}: Quantum
scattering theory for several particle systems. Kluwer Academic
Publishers, 1993.

\bibitem{GraSch}  {\sc G.M.Graf, D.Schenker:} $2$-magnon scattering in the Heisenberg model.
 Ann. Inst. H. Poincar Phys. Theor. 67 (1997), no. 1, 91-107.

\bibitem{Hunziker}	{\sc W. Hunziker:} On the spectra of Schr\"odinger multiparticle
Hamiltonians. Helv. Phys. Acta {\bf 39} (1966), 451-462.

\bibitem{Jorgens}  {\sc K. J\"orgens:} Zur Spektraltheorie der Schr\"odinger Operatoren.
Math. Z.  {\bf 96} (1967), 355-372.

\bibitem{LakMin}  {\sc S.N.Lakaev and Minlos R.A.}:{On bound states of
the cluster operator,} Theor.and Math.Phys.{\bf 39}(1979),
No.1,336-342.

\bibitem{L1}  {\sc S.N.Lakaev}: On an infinite number of three-particle bound
states of a system of quantum lattice particles, Theor.and Math.Phys.{\bf 89}%
(1991),No.1, 1079--1086.


\bibitem{L3}  {\sc S.N.Lakaev}: {The Efimov's Effect of a system of Three
Identical Quantum lattice Particles,} Funkcionalnii analiz i ego
priloj. , {\bf 27}(1993), No.3, pp.15-28, translation in Funct.
Anal.Appl.

\bibitem{LA1}  {\sc S.N.Lakaev, J.I.Abdullaev}: { Finiteness of the discrete
spectrum of the three-particle Schr\"{o}dinger operator on a
lattice,} Theor. Math. Phys. {\bf 111}(1997), 467-479

\bibitem{LS1}  {\sc	 S.N.Lakaev and S.M.Samatov:} On the finiteness of the
discrete spectrum of the Hamiltonian of a system of three
arbitrary particles on a lattice,  Teoret. Mat. Fiz. 129 (2001),
No. 3, 415--431.(Russian)


\bibitem{LA2}  {\sc S.N.Lakaev and J.I.Abdullaev}:{The spectral properties of
the three-particle difference Schr\"{o}dinger operator,}
Funct.Anal. Appl. {\bf 33}(1999), No. 2, 84-88.

\bibitem{LA3}  {\sc	 S.N.Lakaev and Zh.I.Abdullaev:} The spectrum of the
three-particle difference Schr\"{o}dinger operator on a lattice.
Math. Notes, 71 (2002), No. 5-6, 624-633.

\bibitem{AlbLakzM}	{\sc  S.Albeverio, S.N.Lakaev and Z.I.Muminov:}Schr\"{o}dinger
operators on lattices. The Efimov effect and discrete spectrum
asymptotics (to appear in \textit{Annals Inst.Henri
Poincare})

\bibitem{MalMin}  {\sc V.A.Malishev and R.A.Minlos}:Linear infinite-particle operators.Translations of Mathematical
Monographs, 143. American Mathematical Society, Providence, RI,
1995.

\bibitem{Mat}  {\sc D.C.Mattis}:The few-body problem on lattice, Rev.Modern
Phys. {\bf 58}(1986), No. 2, 361-379

\bibitem{Mog}  {\sc A.I.Mogilner}:The problem of a quasi-particles in
solid state physics I n; Application of Self-adjoint Extensions in
Quantum Physics (P.Exner and P.Seba eds.)Lect.Notes Phys.) {\bf
324}, (1998), Springer-Verlag, Berlin

\bibitem{RidSimIV}	{\sc M. Reed} and {\sc B. Simon:} Methods of modern mathematical
physics. III: Scattering theory. Academic Press, N.Y., 1979.

\bibitem{VanWin }  {\sc C. Van Winter}:Theory of finite systems of particles, I.
Mat.-Fys. Skr. Danske Vid.Selsk {\bf 1} (8) (1964), 1-60.

\bibitem{Yaf3}	 {\sc D. R. Yafaev}:
Scattering Theory: Some Old and New Problems.
 (Lecture Notes in Mathematics, Vol. 1735) June 2000  169 pp.  Springer-Verlag, Berlin

\bibitem{Zhislin}  {\sc G. Zhislin}:Discussion of the spectrum
of the Schr\"odinger operator for systems of many particles. Tr. Mosk. Mat. Obs. {\bf 9}
 (1960), 81-128.
\bibitem{Zol} {\.Zoladek, Henryk}:
The essential spectrum of an $N$-particle additive cluster operator,
Teoret. Mat. Fiz. 53 (1982), no. 2, pp.216-226.


\end{thebibliography}
\end{document}